
\input phyzzx.tex
\PHYSREV
\date={1995}

\def\BH{black hole}

\def\beh{\beta_{h}}

\def\gabe{\Gamma_{\beta}}
\def\pano{\par \noindent}
\def\sch{Schwarzschild}
\def\ee{entanglement entropy}

\def\rext{\rho_{ext}}
\def\rint{\rho_{int}}
\def\rh{r_{bh}}
\def\rbo{R_{box}}
\def\text{Tr_{+}}
\def\tint{Tr_{-}}
\def\sent{S_{ent}}

\def\fim{\phi_{-}}
\def\fimp{\phi_{-}^{\prime}}
\def\fip{\phi_{+}}
\def\fipp{\phi_{+}^{\prime}}
\def\omela{\omega_{\lambda}}

\def\erla{R_{\lambda}}
\def\cald{{\cal D}}

\def\calc{{\cal C}}
\def\rh{r_{bh}}

\titlepage
\title{Divergences problem in black hole brick-wall model}
\author{F. Belgiorno\foot{Also Sezione I.N.F.N. di Milano,
 20133 Milano, Italy. E-mail: Belgiorno@vaxmi.mi.infn.it}}
\address{Dipartimento di Fisica, Universit\`a di Milano, 20133 
Milano, Italy}
\author{S. Liberati\foot{Also Sezione I.N.F.N. di Roma, 
 00185 Roma, Italy. E-mail: Liberati@vxrm70.roma1.infn.it}}
\address{Dipartimento di Fisica, Universit\`a ``La Sapienza'' 
di Roma, 00185
Roma, Italy}

\abstract{In this work we review, in the framework of the 
so-called brick wall model, the divergence problem arising in 
the one loop calculations of various thermodynamical 
quantities, 
like entropy, internal energy and heat capacity. Particularly 
we find that, if one imposes that entanglement entropy is equal 
to the Bekenstein-Hawking one, the model gives 
problematic results. 
Then a proposal of solution to the divergence problem is made 
following the zeroth law of black hole mechanics.}

\endpage

\chapter{Introduction}

Recently it was proposed to explain the dynamical origin of 
\BH\ entropy by
identifying it with \ee\ \Ref\FN{V.P.Frolov, 
I.Novikov\journal Phys. Rev.
&D48 (1993)4545.}.\pano
\REF\BFZ{A.O.Barvinsky, V.P.Frolov, A.I.Zelnikov\journal 
Phys.Rev. &D51(95)1741.}
\REF\susug{L.Susskind and J.Uglum\journal Phys.Rev. &D50 
(94)2700.} 
One of the puzzles related with this kind of interpretation is 
the so called
divergence problem. Entanglement entropy is not finite at the 
\BH\ horizon 
[\FN,\BFZ], in order to 
compute that, it is necessary to introduce a kind of 
regularization, by 
imposing an 
ultraviolet cut-off [\FN,\BFZ] either a renormalization of 
gravitational 
coupling constant [\susug] and of constants related to second 
order curvature 
terms \Ref\FS{D.V.Fursaev,S.N.Solodukhin - 
{\em On one-loop renormalization of black hole entropy} - 
Preprint hep-th 9412020.}.\pano
Indeed the first idea  of \ee\ was implicitly proposed by `t 
Hooft in 1985 \Ref\tH{G.`t Hooft\journal Nucl. Phys. &B256 
(1985)727.} applied to his 
``brick wall model"\foot{An extensive study of the divergence 
problem for generic value of the space--time dimension is 
made in 
\Ref\mann{R.B.Mann,L.Tarasov,A.Zelnikov\journal Class. 
Quant. Grav. &9
(1992)1487}.}.
In a certain sense cut-off dependent models [\FN,\BFZ] are up 
to date versions 
of the former. One of the problems `t Hooft proposed in his 
seminal work was
the divergence of not only entropy but also of quantum matter 
contribute to
internal energy of the \BH , which has to be regularized by 
using the same
cut-off one has to introduce for entropy. He found that, fixing 
the cut-off in
order to obtain $\sent=S_{Bek-Haw}=A/4$, one obtains $U={3 
\over {8}}M$. 
So matter
contribution to internal energy appeared to be a very 
consistent fraction of the black hole mass $M$. 
As `t Hooft underlined, this is a signal for a strong 
backreaction effect, not a good aim for a model based
on semiclassical (negligible backreaction) approximation.\pano
We shall show that the same problem is present in BFZ model 
[\BFZ] and that
one  also finds a surprising behaviour of heat-capacity. 

\chapter{Entanglement Entropy and BFZ Model}

In BFZ work entropy is computed from global vacuum density 
matrix by tracing 
over the degrees of freedom of matter outside the \BH. 
In so doing one obtains a mixed state density matrix for matter 
inside the 
\BH. \pano 
BFZ define the global wave function of the \BH\  as:
$$
\Psi=\exp(\Gamma/2) \langle \fim | \exp(-\beta \hat{H}/2) 
| \fip \rangle
\eqn\fdon
$$ 
and
$$
\hat{\rho}=| \Psi \rangle \langle \Psi |
\eqn\den
$$
the related density matrix. Here $|\fip \rangle$ ($|\fim 
\rangle$) are the 
external (internal) states of matter (a massless scalar field for 
simplicity) 
on the \BH\  fixed background.\pano
Tracing over $|\fip \rangle$ gives the internal density matrix:
$$
\eqalign{
\rint(\fimp,\fim)&=\langle \fimp|\hat{\rho}|\fim \rangle 
\cr 
&=\int \cald \fip \Psi^{*}(\fimp,\fip)\Psi(\fim,\fip)\cr
&=\exp(\gabe)\langle \fimp|\exp(-\beta\hat{H})| \fim 
\rangle. \cr} 
\eqn\din
$$
Entanglement entropy associated to this reduced density matrix 
is:
$$
S_{ent}=-Tr_{int}(\rho_{int} \ln \rho_{int})
$$
Here $\gabe$ is a normalization factor fixed in order to obtain $ 
tr \rho=1 $,
but it also corresponds to the 1 loop effective action:
$$
\eqalign{
\gabe &=-\ln \left [ \int \cald \fim \langle \fim | \exp(-
\beta \hat{H}) |\fim
\rangle \right ] \cr
&=-{1\over{2}} \ln \det \left [{{\delta(x-y)}\over{2(\cosh 
\beta
\omega -1)}} \right ]\cr}
\eqn\gam
$$
where $\hat{\omega}$ is the operator associated with the 
frequency of field's
modes.\pano
BFZ calculate the \ee\ as the trace over the internal modes of 
$-\rint \ln \rint $, so their calculation is relative to 
the internal degrees of freedom. Instead, in the common 
definition of \ee, one usually refers to the trace over the 
external degrees of freedom of $- \rext \ln \rext $. In the 
following, we shall show 
that, given the symmetry existing in BFZ study between 
internal 
and external variables, the two definitions of \ee\ coincide.
Besides, given the relation existing between Thermofield 
Dynamics 
and BFZ\rlap,\foot{See for example 
Jacobson\Ref\jaco{T.Jacobson - {\em A  Note on Hartle--
Hawking 
Vacua} - Preprint gr-qc/9407022.}.} we can use the ``external" 
quantities, as defined below, to calculate free energy and 
internal energy relative to the external field.\pano
We recall that BFZ calculation is made in WKB 
approximation.\par
 
From the definitions we have:
$$
\eqalign{
\rext &=\tint | \Psi \rangle \langle \Psi |=\int \cald \fim
\Psi^{*}(\fipp,\fim) \Psi(\fim,\fip)=\cr    
&=\int \cald \fim \langle \fipp| \exp
\left ( - {{\beta \hat{H}}\over{2}} \right ) | \fim \rangle 
\langle \fim | \exp
\left ( -{{\beta \hat{H}}\over{2}} \right ) | \fip \rangle \exp 
(\Gamma_{ext})
\cr
\rint &=\text | \Psi \rangle \langle \Psi |=\int \cald \fip
\Psi^{*}(\fimp,\fip) \Psi(\fip,\fim)=\cr    
&=\int \cald \fip \langle \fimp| \exp
\left ( - {{\beta \hat{H}}\over{2}} \right ) | \fip \rangle 
\langle \fip | \exp
\left (-{{\beta \hat{H}}\over{2}} \right ) | \fim \rangle 
\exp(\Gamma_{int})\cr}
\eqn\den
$$

\section{Symmetry theorem for BFZ model}

We show that for BFZ model it holds an extended version of 
the 
symmetry theorem\Ref\Lib{S.Liberati - Thesis - Universit\`a 
di Roma I ``La 
Sapienza" (1995).}.\pano 
Following Bekenstein path \Ref\Bek{J.D.Bekenstein - {\em Do 
we 
understand black hole entropy ?}, gr-qc/9409015} 
we put
$$
\calc_{\fim,\fip}\equiv \langle \fim | \exp \left ( -{{\beta 
\hat{H}}\over{2}} 
\right ) | \fip 
\rangle.
$$ 
So doing the formulas \den\ become:
$$
\eqalign{
\rext=&(\calc^{\dagger}\calc)_{\fipp,\fip} 
\exp(\Gamma_{ext})\cr
\rint=&(\calc^{*}\calc^{T})_{\fimp,\fim} \exp 
(\Gamma_{int})\cr}
\eqn\rg
$$
From unitarity request for density matrices we obtain:
$$
\eqalign{
\exp(\Gamma_{ext})&={{1}\over{ \int \cald \fip \left.
(\calc^{\dagger}\calc)_{\fipp,\fip} \right |_{\fipp=\fip}}}\cr
\exp(\Gamma_{int})&={{1}\over{ \int \cald \fim \left.
(\calc^{*}\calc^{T})_{\fimp,\fim}\right |_{\fimp=\fim}}}\cr}
\eqn\gfra
$$
so
$$
\eqalign{
\exp(\Gamma_{ext})&={{1}\over{Tr(\calc^{\dagger}\calc)}}\cr
\exp(\Gamma_{int})&={{1}\over{Tr(\calc^{*}\calc^{T})}}\cr}
\eqn\gfrac
$$
By using invariance of the trace under transposition and 
permutation, it is  
easy to see that 
$$
\Gamma_{int}=\Gamma_{ext}.
\eqn\gasim
$$ 
Now $\Gamma_{ext}$ 
can be identified with the product $\beta F$ relative to the 
field degrees of freedom external to the horizon.\pano  
The well--known relations between entropy, internal energy, 
heat capacity and 
free energy 
$$
\eqalign{
S =(\beta \partial_{\beta}-1)\Gamma \cr
U=\partial_{\beta} \Gamma \cr
c=-\beta^{2} \partial_{\beta}^{2} \Gamma \cr}
\eqn\rel
$$
and \gasim\ imply that not only there is no clash between BFZ 
and 
Bombelli, Koul, Lee and Sorkin\Ref\bom{L.Bombelli, 
R.K.Koul, 
J.Lee and R.Sorkin\journal Phys. Rev. &D34(86)373.} definition 
of 
\ee, but moreover that ``symmetry theorem'' is 
generalizable to other thermodynamical quantities, like 
internal energy and
heat capacity.\pano
As a concluding remark of this section, we note that while the 
interpretation 
of the ``internal" entropy \`a la BFZ is clear, it not appears 
obvious which meaning to attribute to the ``internal" free 
energy and internal energy. For example, the internal energy 
for the external fields is 
$$
E^{ext}=Tr(\rho_{ext}  H)
\eqn\dfe
$$
and the equality \gasim\ implies
$$
E^{``int"}=Tr(\rho_{int}  H)=E^{ext}.
\eqn\strange
$$
The Hamiltonian in \strange\ is relative to the external 
degrees 
of freedom: the one relative to the internal ones is $-H$. 
We think that the high symmetry in the variables 
internal--external 
is the reason for this ``extended symmetry theorem''. In other 
words, the property Frolov\Ref\frl{V.Frolov - {\em Black 
Hole Entropy}
- Preprint hep-th/9412211.} calls {\sl duality} is related to this 
very 
special feature of BFZ's model, that in general is not 
implemented. In fact in this case we are defining ``external'' 
and ``internal'' 
thermodynamical quantities on the asymptotically flat regions 
of the 
Einstein--Rosen bridge. Symmetry theorem is just the 
statement of 
the impossibility, for an observer which ``lives''
in one of these regions to discriminate in which of the two he 
is.    
\pano
In what follows, we refer always to the ``external" 
quantities.\par

\section{Effective Action}

Let us calculate $\gabe$, from \gam\ we obtain:
$$
\gabe=\int dx 
\left [ \ln \left ( 2 \sinh {{\beta \hat{\omega}}\over{2}} \right 
) \delta(x-y)
\right]_{\bf{y}=\bf{x}} 
\eqn\ugbfz
$$
where we used the property $\ln \det A= Tr \ln A$.\pano
As BFZ we calculate the expression below by expanding all the 
functions
$\phi(x)$ in terms of eigenfunctions $R_{\lambda}(x)$ of the 
operator
$\hat{\omega}$:
$$
\eqalign{
\phi(x)&=\sum_{\lambda} \phi_{\lambda} 
R_{\lambda}(x)\cr
\hat{\omega}^{2} \erla(x)&=\omela^{2}\erla(x)\cr
\delta(x-y)&=\sum_{\lambda} g^{00} g^{1/2} \erla(x) 
\erla(y) \cr}
\eqn\eigen
$$
where $\sum_{\lambda}$ denotes the sum over all quantum 
numbers, $g^{00}$ is the timelike component of the metric 
tensor and 
$g=\det g_{\mu
\nu}= g^{00}\det g^{ab}$ (a,b,$\ldots = 1,2,3$).\pano
So we obtain: 
$$
\gabe=\int_{2M}^{\rbo} dr {r^{2}\over{(r-
2M)}}\int_{0}^{\infty} 
\sum_{l=0}^{\infty} d\omega (2l+1) R^{2}_{\lambda 
\omega}(r) \gamma (\beta 
\omega) 
\eqn\gammabeta
$$
where
$$
\gamma(\beta \omela)={\beta \over{2}}\omega_{\lambda}+
\ln(1- e^{-\beta \omela})
\eqn\muga
$$
and where $R_{\lambda \omega}(r)$ are the radial 
eigenfunctions. 
We are interested in the behaviour of $\gabe$ near the 
horizon; 
using BFZ result
$$
\sum_{l=0}^{\infty} (2l+1) R^{2}_{\lambda \omega}(r) \sim 
{4\over{\pi}} \omega^{2} {M\over{r-2M}}
\eqn\near
$$
we get 
$$
\gabe \sim {{4M}\over{\pi}} \int_{2M}^{r_{box}} dr
{{r^{3}}\over{(r-2M)^{2}}}\int_{0}^{+\infty} d\omega 
\omega^{2} 
\gamma (\beta \omega)
\eqn\ganear
$$
where $r_{box}$ is the radius of the box in which we have to 
put the \BH\ to
regularize infrared divergences.\pano
To compute the second integral we have to subtract the zero--point 
term from \muga .
We find the following leading term near the horizon:
$$
\gabe=\beta F(\beta)\sim -{{32 \pi^{3} M^{4}}\over{45}} 
{1\over{\beta^{3}}} {1\over{h}}
\eqn\gaho
$$
where the cut--off is defined as $h\equiv Inf(r-2 M)$.\pano
From the free energy \gaho , it is possible to find the other 
thermodynamical
quantities by using \rel\ relations. We obtain: 
$$\eqalign{
S&\sim {{128 \pi^{3} M^{4}}\over{45}} {1\over{\beta^{3}}} 
{1\over{h}}\cr
U&\sim {{32 \pi^{3} M^{4}}\over{15}} {1\over{\beta^{4}}} 
{1\over{h}}\cr
c&\sim {{128 \pi^{3} M^{4}}\over{15}} 
{1\over{\beta^{3}}}{1\over{h}}\cr} 
\eqn\suc
$$
Rewriting the above formulas in terms of a proper distance 
cut--off
$$
\epsilon \sim 2  \sqrt{\rh h} \Leftrightarrow h\sim 
{\epsilon^{2} 
\over{4 \rh}}
\eqn\proper
$$
we find for $F,\ U,\ S\ {\rm and}\ c$ at the Hawking 
temperature
$$\eqalign{
F(\beh)&\simeq -{M \over{720 \pi}} {1\over{\epsilon^{2}}}\cr
S(\beh)&\simeq {2M^2 \over{45}} {1\over{\epsilon^{2}}}\cr
U(\beh)&\simeq  {M \over{240 \pi}} 
{1\over{\epsilon^{2}}}\cr
c(\beh)& \simeq {{4 M^{2}}\over{30 \epsilon^{2}}}\cr}
\eqn\final
$$
The entropy in \final\ is exactly the same than in BFZ.\pano 
\REF\fur{D.V.Fursaev\journal Mod.Phys.Lett. &A10 (95)649.}
\REF\solo{S.N.Solodukhin\journal Phys.Rev. &D51 (95)609.}
\REF\solod{S.N.Solodukhin\journal Phys.Rev. &D51 
(95)618.} 
\REF\fursolo{D.V.Fursaev and S.N.Solodukhin - {\em On 
One-Loop Renormalization 
of Black Hole Entropy} - Preprint hep-th/9412020.}
\REF\froa{V.P.Frolov - Why the Entropy of a Black Hole is 
A/4? - Preprint Alberta-Thy-22-94, gr-qc/9406037.}\par 
Note that \suc\ depends on the implicit standard assumption 
[\BFZ,\susug,\tH,\fur,\solo,\solod,\pano
\fursolo,\froa] of beta--independence 
of the cut--off h appearing in the regularized effective action \gaho. 
We will call in the following ``standard" brick--wall model 
a model in which it holds the above assumption. 
Further discussion is found in the conclusions.

\section{Interpretative Problems}

The divergences appearing in \final\ for the entropy and the other 
thermodynamical quantities requires a renormalization 
scheme or a brick-wall cut-off. The standard position consists in 
identifying the black hole entropy with the leading divergent regularized term:
$$
S_{bh}\equiv S_{radiation,\ leading}. 
\eqn\bwa
$$ 
Our line is to follow the most of papers on the same problem 
[\BFZ,\susug,\tH,\fur,\solo,\solod,\fursolo,\pano \froa] and to 
check which results one can obtain from \bwa\ calculating the 
regularized terms for the other thermodynamical quantities. 
We will work in the framework of 
the brick-wall regularization of the divergences 
[\BFZ,\tH,\froa]; the cut-off in \final\ is the same for all the 
thermodynamical quantities: we have to use the same value of the 
cut--off, fixed by eqn. \bwa, for all of them.\par
The identification of Bekenstein-Hawking entropy with the 
entanglement one of course generates a problem of 
interpretation of classical (tree level) entropy due to gravity in 
the path-integral approach. The first aim of entanglement approach is 
to explain as dynamical matter entropy
all \BH\ entropy. The matter leading term is not a new one-
loop contribution to be added to the tree level one. 
So it appears as a necessary complement of this
program a clear explanation for ignoring the presence of the 
tree level contribution of gravity. As a matter of fact in 
literature this problem appears 
to be often ignored or gone around. We can quote in this sense 
only a work 
by Jacobson \Ref\Jac{T. Jacobson - {\em Black hole entropy and 
induced 
gravity} -Preprint gr-qc/9404039} and an alternative proposal by 
Frolov [\froa].\pano 
The interpretative problem\foot{`t Hooft, as a matter of facts, 
seems aware of the 
interpretative problem; his insight is that also the mass of the 
black hole should be entirely due to the radiation 
\Ref\thosc{G.'t Hooft -
Scattering Matrix for a Quantized Black Hole - in {\sl Black 
Hole Physics}, 
V. De Sabbata and Z. Zhang eds, Kluwer (1992).} 
Anyway, the problem of how implement an identification of 
black hole 
quantities and radiation ones, is left open.}
 is worse for the other thermodynamical 
quantities.\pano
 For the internal energy,
we will find that it seems impossible the identification 
of the brick-wall value with the tree level one; 
on the other hand, it does not seem 
possible to understand the radiation term as a 
perturbative contribution to the black hole tree level one. In 
the second case, the underlying idea is that geometry 
(\BH\ internal energy $M$) isn't induced by linearized matter 
fields in thermal equilibrium with the black hole.
A similar situation is found for heat capacity.\pano
We stress that self-consistency check of \bwa\ imposes to 
compare tree level gravitational values with the 
1-loop matter ones. 

\section{Free Energy and Internal Energy}

The cut-off fixing necessary to obtain the required value

$S_{ent}=S_{Bek-Haw}=A/4$ is:
$$
\epsilon^{2}={{1}\over{90 \pi}}
\eqn\cuto
$$
this brings to the following values for free energy and internal 
energy:
$$\eqalign{
F&=-{{1}\over{8}}M\cr
U&={{3}\over{8}}M.\cr}
\eqn\udef
$$
The results in \udef\ are the same obtained by t'Hooft in his 
pioneering 
paper [\tH] and exactly the same are found if one calculates $U$ 
and $F$ with 
heat kernel expansion truncated to the first De Witt 
coefficient in the optical metric\foot{See the following 
section.}.\par

\section{Heat Capacity}

We now want to show the behaviour of \BH\ heat capacity in 
brick wall model.\pano
For heat capacity, imposing again the cut-off value \cuto , one 
obtains from
\final\ 
\foot{Also in this case, it is possible to obtain the same value 
from `t Hooft results [\tH] with 
a simple computation using his internal energy.}:
$$
c=+12\pi M^{2}
\eqn\cfin
$$
It is important to note that this is a positive value and in 
module bigger
than the classical well-known result:
$$
c_{class}=-8 \pi M^{2}
\eqn\ccla
$$
So if we accept the brick-wall model plus entanglement entropy 
frame as
dynamical explanation of \BH\ entropy we find, in the most 
naive 
interpretation of \cfin, that \BH s are 
stabilized by one loop contribution of matter.
On the other side \cfin\ is different from 
its classical counterparts not only 
for a numerical difference but also because they describe 
completely different thermodynamical objects. \pano
The same value \cfin\ can also be found by using well-known 
results 
\Ref\DK{J.S.Dowker, G. Kennedy\journal Jou. Phys. &A11 
(78)895.} 
for thermodynamical quantities for scalar field confined in 
spatial
cavity in a static space-time at finite temperature. In [\DK] a 
high temperature
expansion, in terms of the De Witt coefficients, is performed.
From [\DK] the main contribution to entropy is given by:
$$
S= {{2 \pi^{2}}\over{45}} {{1}\over{\beta^{3}}}\cdot \bar{c}_{0}
$$
Here $\bar{c}_{0}$ is the first De Witt coefficient in optical 
metric 
$g^{opt}_{\mu \nu}\equiv {\bar{g}}_{\mu \nu}=g_{\mu 
\nu}/g_{00}$ which in our 
case is:
$$
\eqalign{
\bar{c}_{0}&=\int d^{3}x \sqrt{\bar{g}}\cr
&= 4 \pi \int dr r^{2} \cdot{{1}\over{g^{2}_{00}(r)}}\cr
&= 4 \pi \int_{2M}^{R} dr {{r^{4}}\over{(r-2M)^{2}}}\cr
& \approx 4 \pi \cdot {{(2M)^{4}}\over{h}}\cr}
\eqn\cM
$$
So at the Hawking temperature the entropy for the \sch\ 
\BH\ is:
$$
S = {{M}\over{180}} \cdot {{1}\over{h}}
\eqn\Sm
$$
To get $S=A/4=4\pi M^{2}$ we have to fix (as `t Hooft) $h 
\equiv {1\over{720 \pi M}}$.
We can now calculate heat capacity. We find:
$$
\eqalign{
c&= {{2 \pi^{2}}\over{15}} \cdot {{1}\over{\beta^{3}}}\cdot 
\bar{c}_{0} =\cr
&= {{M}\over{60}}\cdot {{1}\over{h}}\cr}
\eqn\hcdk
$$
Here we have used \cM.\pano
Introducing in \hcdk\ the former cut-off fixing, we finally 
obtain:
$$
c={{M}\over{60}} \cdot 720 \pi M = 12 \pi M^{2}.
\eqn\hcfin
$$

\chapter{Conclusions and Perspectives}

Our analysis suggests that brick wall model interpretation of 
\BH\ entropy brings to problematic results 
for internal energy and heat capacity. 
The internal energy, as `t Hooft remarked [\tH], 
is of the same order of magnitude of \BH\ mass: at this point 
one must question the applicability of the assumption of the 
negligible back-reaction.
Even if one passes over this problem, we still find that the one 
loop contribute of matter to \BH\ heat capacity is positive and 
so it would stabilize the \BH\ . But we believe it is an 
inconsistent result because 
quantum correction is for the heat capacity bigger than its 
background counterpart; it could be more plausible 
if we would have in the gravitational action quadratic terms 
in curvature tensor, but this is not our case.\par 
Our results can be interpreted as a ``warning bell'' of a structural 
problem embedded in the standard brick--wall approach to \BH\ 
thermodynamics. Perhaps it is due to the fact that one ignores 
the back-reaction of matter field 
on the gravitational background. 
Maybe that relaxing the standard assumption of 
beta-independence of the cut--off h appearing in the regularized 
effective action \gaho\ one eventually gets 
a consistency of all thermodynamical quantities.\par
Our proposal is to review critically\Ref\fra{F.Belgiorno - Ph.D. Thesis- 
Universit\`a degli Studi di Milano (1995).} the key idea 
underlying the usual approach, according to which, in order to 
take 
the $\beta$-derivatives 
necessary to compute from the partition function the various 
thermodynamical 
quantities we need to displace slightly from $\beta_{h}$, 
introducing a conical 
singularity in the manifold. 
The above approach understands that, as it is made in 
``common" 
manifolds not characterised by the Hawking effect, it is possible 
to give 
to the parameter $\beta$ the meaning of an allowed physical 
equilibrium 
temperature for the quantum fields living on the manifold. So 
all the 
physical quantities are calculable from the partition function by 
mean of 
$\beta$-derivatives. \pano
The case of a manifold characterised by the Hawking effect 
is in the conical approach treated on the same foot: in order to 
study 
the equilibrium thermodynamics at the Hawking temperature, 
we have to 
take the $\beta$ derivatives and then to put $\beta=\beta_{h}$. 
This is equivalent to introduce a displacement in the manifold 
from its 
natural period.\par \noindent
Our ansatz [\fra] is that equilibrium thermodynamics in a black 
hole manifold 
requires a different approach that takes its stand essentially 
from a 
literal interpretation of the zeroth law of black hole 
mechanics \Ref\bard{B.Bardeen,B.Carter and 
S.W.Hawking\journal Commun.Math.Phys. 
&31 (73)161.}.\pano
Indeed, we know that for manifolds with a bifurcate Killing 
horizon the equivalent of the temperature concept is given by 
the surface 
gravity $k$; they are related by the formula
$$
T={k\over{2 \pi}}.
\eqn\surf
$$ 
So, if we allow the temperature, or equivalently, its inverse 
$\beta$, to vary freely (as it is true in the canonical ensemble, 
in which all the above calculation are made), then we have to 
think the geometrical parameters entering in the surface 
gravity 
as functions of the temperature by mean of \surf. To 
be 
more clear, the equation defining the proper period of the 
manifold
$$
\beta={{2 \pi}\over{k}}
\eqn\ans
$$
has to be considered a constraint equation for the geometrical 
parameters appearing in \ans. 
In the case of the Schwarzschild black-hole 
$\beta_{h}$ is the Euclidean time period to be 
naturally selected with the aim of avoiding the conical 
singularity. 
The equilibrium thermodynamics of the manifold is such that 
one 
has to adjust the relevant geometric parameters in order to 
match 
the generic $\beta$ and the surface gravity $k$ associated with 
the horizon: 
given the relation 
$$
\beta={{2 \pi} \over {k}}=8 \pi M
\eqn\ha
$$
we have to substitute in the metric 
$$
M={\beta \over{8 \pi}}.
\eqn\bel
$$
We note that, to make the above match, the metric becomes 
$\beta$--dependent.\par
The ansatz is coherent with the perturbative expansion of the 
path integral of the finite temperature quantum field theory: 
the starting point in the path--integral calculation of 
the partition function is a tree level approximation for the 
gravitational part; the classical solution we are interested in 
is the \sch\ solution. This solution is 
smooth and gives the relation \bel\ between its period in 
Euclidean time and the black hole radius.\pano 
All the matter field contributions in the linearized theory are 
perturbations of the tree level ones.\par 
Besides, as far as the matter field backreaction is not included, 
matter fields are not able to modify the link between geometry 
and 
thermodynamics given by \surf.\par
We stress that, in this way, 
all the diseases due to introduction of the conical defect 
are eliminated.\pano
\REF\gross{D.J.Gross and M.J.Perry, L.G.Yaffe\journal Phys. 
Rev. 
&D25 (82)330.}
In support to this approach we cite analogous statements in 
[\gross, 
\froa]. We think that it incorporates the tree-level back reaction 
phenomenon, and on the other
hand, it should make the one-loop contributions to be a small 
correction to the 
tree level ones; particularly, the radiation entropy should be 
only a 
small perturbation w.r.t. the big 
Bekenstein-Hawking entropy.\par
We mean to come back on these topics in a next publication 
\Ref\bellib{F.Belgiorno, S.Liberati and M.Martellini - Work in 
progress.}.

\ACK{The authors wish to thank M. Martellini and K. Yoshida 
for extensive
discussions and some constructive advice, and the referee for 
suggestions and critical remarks. S. Liberati thanks V. 
Ferrari for illuminating remarks.}

\endpage\refout
\end